\pdfoutput=1

\documentclass{sig-alternate}

\include{goodmathstuff.tex}
\usepackage{array,multirow}  

\usepackage[table,x11names]{xcolor}  

\newcommand{\highlightrow}{}  
\begin{document}
%
\conferenceinfo{CISR}{'16 Oak Ridge, Tennessee USA}

\title{HackAttack: Game-Theoretic Analysis\\
of Realistic Cyber Conflicts
}

%
%
%
%
%

\numberofauthors{6} 
%
\author{
%
%
\alignauthor
Erik M.~Ferragut\\
       \affaddr{Oak Ridge National Laboratory}\\
       \affaddr{Oak Ridge, Tennessee}\\
\alignauthor
Andrew C.~Brady\\
       \affaddr{Jefferson Middle School}\\ 
       \affaddr{Oak Ridge, Tennessee}\\
\alignauthor
Ethan J.~Brady\\
       \affaddr{Oak Ridge High School}\\ 
       \affaddr{Oak Ridge, Tennessee}\\
\and
\alignauthor
Jacob M.~Ferragut\\
       \affaddr{Oak Ridge High School}\\ 
       \affaddr{Oak Ridge, Tennessee}\\
\alignauthor
Nathan M.~Ferragut\\
       \affaddr{Oak Ridge High School}\\ 
       \affaddr{Oak Ridge, Tennessee}\\
\alignauthor
Max C.~Wildgruber\\
       \affaddr{Oak Ridge High School}\\ 
       \affaddr{Oak Ridge, Tennessee}\\
}

\date{30 October 2015}

\maketitle
\begin{abstract}
  Game theory is appropriate for studying cyber conflict because it
  allows for an intelligent and goal-driven adversary. Applications of
  game theory have led to a number of results regarding optimal attack
  and defense strategies.
  However, the overwhelming majority of applications explore
  overly simplistic games, often ones in which each participant's
  actions are visible to every other participant. These
  simplifications strip away the fundamental properties of real cyber
  conflicts: probabilistic alerting, hidden actions, unknown opponent
  capabilities.

  In this paper, we demonstrate that it is possible to analyze a more
  realistic game, one in which different resources have different
  weaknesses, players have different exploits, and moves occur in
  secrecy, but they can be detected. Certainly, more advanced and
  complex games are possible, but the game presented here is more
  realistic than any other game we know of in the scientific
  literature. While optimal strategies can be found for simpler games
  using calculus, case-by-case analysis, or, for stochastic games,
  Q-learning, our more complex game is more naturally analyzed using the same
  methods used to study other complex games, such as checkers and
  chess. We define a simple evaluation function and employ 
  multi-step searches to create strategies. We show that such scenarios can be
  analyzed, and find that in cases of extreme uncertainty, it is often
  better to ignore one's opponent's possible moves. Furthermore, we
  show that a simple evaluation function in a complex game can lead to
  interesting and nuanced strategies.
\end{abstract}




\vfill 
\eject
\section{Introduction}
\label{sec:intro}
Computers connected to the Internet are potential targets for criminals, botnets, and national cyber capabilities. Cyber conflicts result from competing interests struggling to control the same assets. If one side of the conflict is manually assessing the situation, reasoning about the risks, making decisions, and executing attacks it will likely be overwhelmed if the other side adopts an automated approach in which algorithms are used to make every choice. The speed of human-in-the-loop decision making is likely to be several orders of magnitude slower than a well tuned algorithm. The future of cyber offense and defense undoubtedly depends on the development of advanced, scalable, and effective algorithms for decision-making in real-time. Moreover, the organization that first masters this technology will likely obtain a decisive advantage over their adversaries. 

A number of challenges remain before effective cyber conflict decision algorithms can be developed. Current methods lack a framework with which to automate reactions in realistic scenarios. In particular, what are relevant assets and capabilities? How can they be used? What is the relative benefit of each choice? And, given answers to these questions, how can the best actions be chosen and coordinated? These questions are the province of game theory.

In this paper, we take a game-theoretic approach to developing a prototype for intelligent, automated attack and defense. Previous work in this direction has analyzed simplified abstractions of cyber situations. We go further in this work by (1) creating a new game, HackAttack, that can be used to model larger, more complex conflicts than previous work, and (2) comparing and analyzing a number of strategies for HackAttack. These strategies use tree searches to maximize an evaluation function. We show that a simple evaluation function in a complex game can lead to interesting and nuanced strategies. This provides a promising path forward for developing a method for intelligent cyber conflict automation.

In Section~\ref{sec:background}, we review the extensive literature on applications of game theory to cyber security, and emphasize how this work extends it. In Section~\ref{sec:hackattack}, we introduce our new game, HackAttack, and describe its rules and concepts in detail. To play the game, the player strategies are created using tree searches and an evaluation function, which are described in Section~\ref{sec:strategies}. Strategies are analyzed in simulated games, as described in Section~\ref{sec:results}. In Section~\ref{sec:discussion} we summarize the lessons learned and implications for future work in regards to searching trees to select moves in cyber conflicts. We summarize our conclusions in Section~\ref{sec:conc}.

\section{Background}
\label{sec:background}
An extensive literature now exists that analyzes cyber security using
game theory. However, there has been very little work analyzing realistic
games. Games can be categorized by various independent dimensions:
static or dynamic, complete or incomplete, perfect
or imperfect. Static games are games in which the order of the moves
do not matter; these are generally viewed as games where the players
move simultaneously. Non-static, or dynamic, games are more realistic
for modeling cyber conflict. In a game of complete information, each
player knows the payoff function of each of the other players, whereas
this is not the case for games of incomplete information. Finally, in
games of perfect information, each player observes (without error) all
actions taken and their outcomes. A notable feature of cyber conflict
is that each agent attempts to remain undetected; games where actions
or outcomes may be unobserved or observed with error or uncertainty
are called games of imperfect information. In this paper, we develop a
dynamic game of complete but imperfect information. For useful surveys
that describe the literature along this taxonomy,
see~\cite{liang2013game} and~\cite{roy2010survey}.


A number of researchers have studied (or primarily focused on) static
games, such as~\cite{chen2009game,grossklags2009uncertainty,he2012game,liu2006bayesian}. These
games have the significant advantage that they can be analyzed (e.g.,
Nash equilibria can be computed) using existing methods. In some cases
(e.g.,~\cite{nguyen2009security}) the repeated game is studied, which
adds some realism, but fails to capture the essence of cyber attacks
where a sequence of small gains may lead to a dominant position.


Among dynamic games, it is common to assume perfect information where
each player can observe the actions of the others. For example,
in~\cite{luo2010game, nguyen2009security, patcha2004game, 
  sagduyu2011jamming, shamma2005dynamic}, attacks are always
detected. This renders the results inapplicable to real cyber
conflicts, since cyber attacks are often undetected.


Almost all work involving dynamic games with imperfect information
focuses on so-called stochastic games. In these games
(e.g.,~\cite{alpcan2006intrusion, lye2005game, nguyen2009stochastic,
  sallhammar2006stochastic, xiaolin2008markov}), there is a finite
state space, and transition probabilities are determined by the
simultaneous move choices made by the players. Special methods have
been developed to find optimal strategies for these games, such as the
Q-Learning method~\cite{hu2003nash} and methods from competitive
Markov Decision Processes~\cite{filar2012competitive}. However, these
methods scale poorly with problem size, so we adopt different methods
in this work.

The survey~\cite{roy2010survey} describes the lack of realism in cyber
games as a key issue in previous research.
\begin{quote}
  Some of the limitations of the present research are: (a) Current
  stochastic game models only consider perfect information and
  assume that the defender is always able to detect attacks; (b)
  Current stochastic game models assume that the state transition
  probabilities are fixed before the game starts \ldots; 
  (c) Current game models assume that the players' actions
  are synchronous, which is not always realistic; (d) Most models are
  not scalable with the size and complexity of the system under
  consideration.
\end{quote}
In this work, we address the first three of these concerns.

One issue all previous games have is that payoff functions must be
assigned to certain outcomes. These numbers typically must be chosen
to represent some intuitive notion of the benefit of each outcome, but
the payoffs, being very difficult to validate, must generally be
accepted as notional. In our game, we instead define a win condition. Namely,
the attacker who owns the most assets at the end of a given number of
time steps is the winner. This view of the payoff directly correlates
with the real-world goals of attackers and is therefore more
defensible.

Another key distinction is that {\em all} previous work we are aware
of analyzes an attacker-defender situation. We analyze a two-attacker
situation where two adversaries are attempting to control a common
pool of resources. This allows for a symmetry in the strategies and
avoids having to construct two different kinds of strategies (i.e.,
attackers and defenders). Consequently, the analysis and comparison of
strategies is simplified. Also, adding defenders at a future time will
be possible by augmenting the game with ownership and defensive moves
(e.g., clean-booting).

Furthermore, the complexity of our game makes it very difficult to use
the strategy optimization methods popular in the literature. Instead,
we adopt the strategy analysis methods used in games such as
chess. These include using an evaluation function and performing
$k$-ply searches, as described in~\cite{levy2009computer,shannon1950programming}.

\section{HackAttack}
\label{sec:hackattack}
We developed the game, HackAttack, specifically for this research to simplify cyber conflict into an analyzable game while avoiding oversimplification. 
In HackAttack, players battle over a pool of resources, similar to how two actors might attempt to grow their botnets. 
The winner is the player who has the most computers after a set time frame. 
The game is turn based in a fixed player order, and players take one action each turn for each resource they control. The resources are a collection of ``neutral'' computers accessible from the Internet. 

The rules of this game apply to any number of players, and the game has been played with up to five players to test the game mechanics. However, for the purposes of this analysis, only two players were used. 

{\em Computers and Accounts.}
The game is played with $5p$ computers where $p$ is the number of players.
For example, in a two-player game, there are 10 available computers.
Each player has a starting computer on which they have one account.
Accounts represent the presence a player has on a computer. Each player begins with one account on their starting computer
More accounts make it easier to remove the other player off a computer and to resist being removed off a computer.
We say that a player {\em controls} a computer if they have at least one account on it.
The most accounts any player can have on any machine is four. 
Multiple players can have a presence on a computer, with each of them having up to four accounts on it at the same time. 

{\em Exploits.}
Exploits are named by the operating system (OS) they target and their power level.  At the start of the game, computers are randomly, but equiprobably designated a certain OS, indicated by four types, 0 through 3. (These can be thought to stand for Linux, Mac, Solaris, and Windows, for example, but the specific labels have no bearing on the game.) 

The power levels for an exploit range from 0 to 14. The chance of getting an exploit with power level $n$ is $2^{-(n+1)}$. Therefore an exploit with power 0 has a 50\% chance of being found, an exploit with power 1 a 25\% chance, all the way to exploits with power 14, which has about a 0.003\% chance of being found\footnote{Technically, the probabilities are then normalized to sum to 1 by dividing by (100 - 0.003)\%.}. Each player starts out with four exploits, and there is a 1/6 chance of gaining a random exploit each round. 

{\em Vulnerabilities and Patches.}
Computers at the start of the game are patched against three exploits, unknown to the players, representing the antivirus software the host user already has. These patches are chosen with the same probability as the exploits for players; weaker ones are more common.

{\em Actions.}
On their turn, each player gets to assign one action to each computer they control.
Actions are types of attacking, defending, and scouting. These moves are summarized in Table~\ref{tab:moves}.
	
\begin{table*}
\begin{center}
\begin{tabular}{|l|l|r|r|}
\hline 
\highlightrow \multicolumn{4}{|c|}{Attack} \\
\hline 
{\tt Hacking} & Use an exploit from one computer to gain an account on another &  20\% P.D. & Remote \\
\hline 
\highlightrow \multicolumn{4}{|c|}{Defense} \\
\hline 
{\tt Backdooring} & Add more ways to control a computer & 15\% P.D.  & Local \\
{\tt Cleaning} & Remove other player's accounts from a computer & 100\% P.D. & Local \\
{\tt Patching} & Block all future uses of an exploit on a computer & 25\% P.D.  & Local \\
\hline 
\highlightrow \multicolumn{4}{|c|}{Scouting} \\
\hline 
{\tt Reconning} & Identify a computer's OS and vulnerabilities & 5\% P.D. & Remote \\
{\tt Scanning} & Use an account on a computer to find the presence of other players & 30\% P.D. 
& Local \\
\hline
\end{tabular}
\caption{Summary of available moves in HackAttack. PD = Probability of Detection. {\tt Hacking} and {\tt recon} affect a targeted computer from a controlled computer. The other actions affect the controlled computer itself.\label{tab:moves}}
\end{center}
\end{table*}

Attacking, in the game called {\tt hacking}, is using an exploit to gain one account on a computer.
An exploit for a specific OS only works for that one OS.  
Exploits are the only form of attack, and using different exploits is the only way to change how you attack. If the {\tt hacking} succeeds, the one carrying out the hack gains an account on the target machine.  
A successful hack puts one account on the computer targeted, up to the maximum number of accounts. 
If the {\tt hacking} fails, the attacker learns no additional information about the target. 

Defense is accomplished by {\tt patching} to decrease the likelihood of an opponent successfully attacking your computer, {\tt cleaning} to remove your opponent from a computer, or {\tt backdooring} to making it harder for you to be removed off a computer. 
{\tt Backdooring} puts one more account on a computer that you own. 
{\tt Cleaning} removes enemy accounts off a computer equal to the accounts you have on that computer.  Removing a player from a computer is important because only computers with at least one player account on it will count toward a player's final computer total. After {\tt cleaning}, you know how many enemy accounts were removed.  The different outcomes of this have varying meanings, and certain results make it unclear if the opponent is still on the computer. {\tt Cleaning} is especially useful after you detect an opponent using that machine.
 {\tt Patching} makes a computer you own permanently safe against an exploit you choose. Since you only know about the exploits you own and use to hack, you can only patch exploits you own. 
One downside of {\tt patching} is that if you patch an exploit and are later removed from the computer, then you cannot use the exploit you patched to get back on that computer. 

The scouting functions, {\tt scanning} and {\tt reconning}, reveal information about a computer. {\tt Scanning} tells you who else, if anyone, is on the targeted computer you own and the amount of accounts they have on it.  It may be used before {\tt cleaning} to know if your {\tt clean} will entirely remove your opponent. Unlike {\tt cleaning}, {\tt scanning} tells you definitively if someone else is on a computer and if they will remain after a {\tt clean}.  {\tt Reconning} one computer from a controlled computer tells you the OS of the targeted computer and what exploits you have, if any, that can {\tt hack} it, but it does not indicate whether it is already occupied. This action is used to set up a {\tt hack} on the following turn because it can almost guarantee the {\tt hack}'s success.  If you are {\tt cleaned} off that computer and want to get back on, you would not need to {\tt recon} again because you would remember the computer's OS.

{\em Detection.}
It is important to note that each player only observes another player's action if it is ``detected''. Otherwise, they are aware only of their own actions and the results of those actions. Every action, therefore, has a detection probability based on how much it interacts with the computer.  {\tt Patching} has a 25\% detection rate, {\tt hacking} 20\%, {\tt backdooring} 15\%, {\tt reconning} 5\%, {\tt scanning} 30\%, and {\tt cleaning} 100\%. One important thing to note is that you can be detected by a player on the targeted machine, and, in the case of {\tt hacking} and {\tt reconning}, on the acting computer.

{\em Win Condition.}
After 20 rounds, the player controlling the greatest number of computers is the winner, or a tie is declared if each player controls the same number of machines. A game is stopped early if one player is {\tt cleaned} off of every computer. 
Having at least one account on a computer makes it contribute to your total number of machines controlled at the end of the twenty rounds, and both players can have the same computer count toward their totals.

\section{Strategies}
\label{sec:strategies}

A strategy for a game is a function that takes as input a player's situation and returns that player's move(s). There are many ways to develop strategies. A common approach in the cyber security game theory literature is to explicitly write down the objective function and then directly derive optimal solutions. This approach works well only in the simplest games, such as static matrix games. For more complex games, such as stochastic Markov games, alternative methods (e.g., Q-Learning) can be used~\cite{alpcan2006intrusion}. However, the most complex games must adopt methods similar to those used to develop strategies for checkers and chess. These approaches involve the development and use of {\em evaluation functions} to score the desirability of any possible state of the game. The evaluation function is a formalization of expert intuition on what constitutes a good position. For example, in the game of chess, it may include points for material, controlling the center of the board, having freedom to move, and so on. The evaluation function is then used in a multiple look-ahead tree search to select the best move to play.

\begin{figure}[h!tbp]
\begin{center}
\includegraphics[width=2.8in]{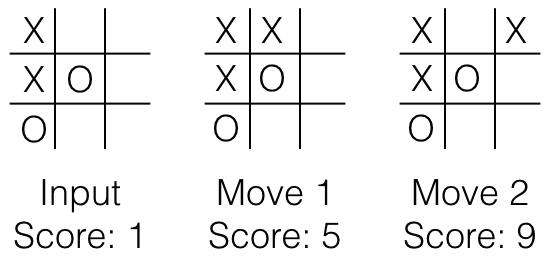}
\caption{An evaluation function for scoring the tic-tac-toe board for player X could be defined by granting 4 points for each open two-in-a-row that X has, subtracting 1 for each one O has, subtracting 1 point if O occupies the center, and adding 3 points for each corner occupied by X.
According to this evaluation function, X would prefer Move 2 to Move 1 since it has a higher score.\label{fig:tictactoe}}
\end{center}
\end{figure}

As a simple example, a possible tic-tac-toe evaluation function is described in Figure~\ref{fig:tictactoe}. Such an evaluation function can then be used to consider what-if scenarios and select a move. For example, Figure~\ref{fig:tictactoe} shows a board and two possible moves that have been scored according to this evaluation function. This approach suggests that the second move is preferable to the first. 

In practice, however, it can be very difficult to engineer an evaluation function that fully captures the intricacies of the game. The value of a game state depends on the possible responses by the opponents. 
For example, points could be added if a good move could be possible on the next turn and subtracted if the opponent can make a good move. 
The combinatorial possibilities make such an evaluation function difficult to define and harder to tune.
Rather than attempt to construct a highly complex evaluation function, an alternative method is to ``look several moves ahead'' by searching a tree of subsequent possible.

\begin{figure}[h!tbp]
\includegraphics[width=3.3in]{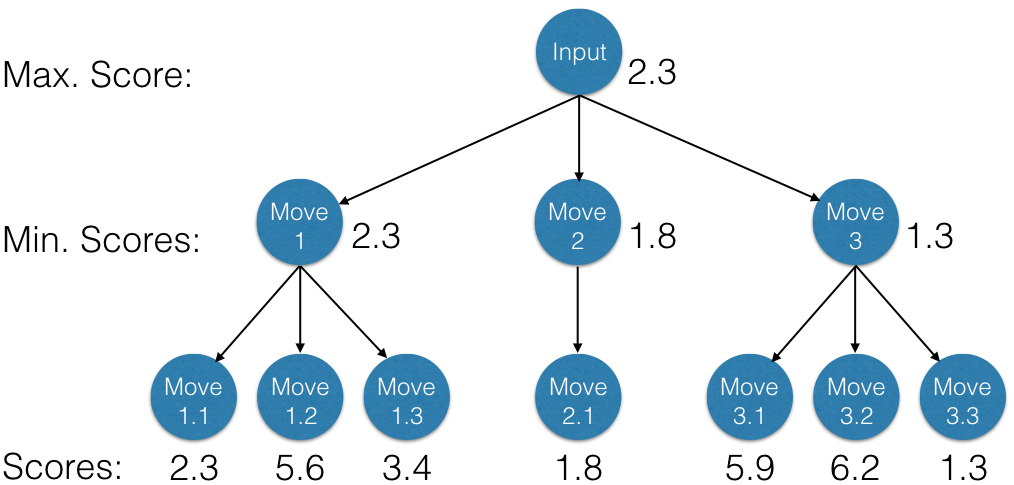}
\caption{A 2-ply tree search explores all moves, starting from an input state, to create the middle layer of the tree. From each of those states, all possible opponent moves are considered. The leaf nodes are then scored by the evaluation function. The middle layers are assigned a score, not from the evaluation function applied to their states, but as the minimum of their children's scores. The root node is given the maximum of its children's scores. In this case the best move would be Move 1, since it was the source of the high score.\label{fig:minmaxtree}}
\end{figure}

As shown in Figure~\ref{fig:minmaxtree}, a tree is constructed top-down starting with the current board configuration as the root. For each possible move, a child node is created. This makes up the {\em 1-ply} tree. Then, at each leaf, each opponent move is added as a child, making up the {\em 2-ply} tree. This process continues to the desired depth. At the bottom, each board configuration is scored according to the evaluation function. These scores are then rolled up as follows. If the last move represents the player's move then, under the assumption that the player will try to improve the board, the maximum of the childs' scores are assigned to the parent. On the other hand, if the last move represents the opponent's move then the minimum is used. This process is repeated up the tree, alternating between min and max. The score given to the root is the largest value among its children (which may occur multiple times). One of these highest-scoring moves is then selected. This is the {\em $k$-ply strategy} for the given evaluation function. Evaluation functions and $k$-ply search trees have been used to analyze a wide variety of games~\cite{levy2009computer}. 

The basic $k$-ply strategy assumes that all possible moves are known and deterministic. It also assumes that each player gets one move. In HackAttack, results are probabilistic and the number of moves a player gets next turn depends on the outcome of their move this turn. Furthermore, each player has only limited knowledge of the other players' knowledge, accounts, and exploits. This lack of knowledge makes it very difficult to consider the opponent's best move; that would require the computationally intractable problem of branching over all possible assets the opponent may have and averaging the results according to the probabilities. Consequently, we adopted some simplifying assumptions and explored two different kinds of $k$-ply search: Random-Response and No-Response.

In a Random-Response $k$-ply search, the opponent's move is modeled as being randomly chosen from all possible moves. The probability of each move is weighed by the probability that the opponent would have the resources to carry out the move, according to the player's knowledge. To this end, each player's knowledge of the other player had to be tracked, which is described below.

In a No-Response $k$-ply search, the opponent moves are not considered. Each level of the tree contains the next possible moves for the player. This approach has the disadvantage of not being able to account for enemy choices. However, in HackAttack, often very little is known about the enemy and their resources, which makes it difficult to respond meaningfully. In this tree search, each node is assigned the maximum score of its children nodes. The main advantage of the No-Response $k$-ply search is that it is possible to develop more elaborate (i.e., multi-step) plans.
The name No-Response indicates that it plans ahead as if no response will be brought against its actions, but the strategy itself can still respond to its opponent. 
For example, if a successful {\tt hack} is detected, it could decide to {\tt clean} that computer in response.
 
In HackAttack, a player is allowed one move for each computer on which they have at least one account. This means that a player with, say, 5 accounts, would have an enormous number of possible moves. This combinatorial explosion slows the tree search and makes it impractical for analysis. To avoid this issue, we consider a single computer at a time. A tree is created for the moves from that computer, and the best move is chosen. The result of that move is then used as the starting point of a tree search for the next computer. This approach is less thorough and may miss some possibilities for coordinating actions, but it turns a multiplicative work factor into an additive one. 

The information at each node in the search tree is a knowledge structure. It contains four tables of probabilities indicating the player's degree of belief in various claims. 
The probabilities are updated using Bayes's Theorem when the player observes something (e.g., the result of their action or the detection of an opponent's action).
The first table gives the probability of each host having a particular OS. The second table tracks the probability of each machine being patched against each exploit. The third table provides the probability of player $p$ having $r$ accounts on computer $c$. The last table contains the probability that player $p$ has each possible exploit. 

The use of a $k$-ply tree search requires the specification of an evaluation function. For the purposes of this research, we used a ``net-computers'' evaluation function, which is the number of computers the player has accounts on minus the expected (in the probabilistic sense) number of computers the opponent has. (The expected number is used because the exact number is not known to the players.) 


\section{Results}
\label{sec:results}




Each of four strategies were tested against each of the others with each strategy going first or second in ten games for each configuration.
The four strategies are (1) Random, which picks a legal move uniformly at random, (2) One-Step, is a 1-ply search that looks at only the first move, 
(3) Random-Response, which looks two moves ahead and models the opponent's move as Random, and 
(4) No-Response, which considers making two of its own moves and ignores the opponent's response.
All four methods use the net-computers evaluation function.
Table~\ref{tab:results} summarizes these pairwise results.

\begin{table}[h!tb]
\begin{center}
\begin{tabular}{|c|c|c|c|c|} \hline
\highlightrow
 Strategy for P1 & Strategy for P2 & \multicolumn{2}{c|}{Wins} & Ties \\ \cline{3-4}
\highlightrow & & P1 & P2 & \\ \hline
Random & {\bf One-Step} & 3 & 6 & 1 \\
{\bf One-Step} & Random &5 & 2 & 3 \\
Random & {\bf Rand-Response} & 0 & 9 & 1 \\
{\bf Rand-Response} & Random & 8 & 0 & 2 \\
Random & {\bf No-Response} & 2 & 8 & 0 \\
{\bf No-Response} & Random & 10 & 0 & 0 \\
One-Step & Rand-Response & 4 & 6 & 0 \\
Rand-Response & One-Step & 5 & 5 & 0 \\
One-Step & {\bf No-Response} & 2 & 7 & 1 \\
{\bf No-Response} & One-Step & 7 & 3 & 0 \\
Rand-Response & {\bf No-Response} &  3 & 7 & 0 \\
{\bf No-Response} & Rand-Response & 9 & 1 & 0 \\
\hline
\end{tabular}
\caption{Results from comparing each pair of strategies in ten games each. The winning strategy is shown in bold text.\label{tab:results}}
\end{center}
\end{table}

The results show that Random is the worst strategy and that No-Response is the best against the strategies tested. 
One-Step and Random-Response performed comparably and evenly split wins when pitted against each other. They both lost approximately the same number of games to No-Response. 
However, the Random-Response strategy won more often against Random than One-Step did.
In summary, the strategy rankings are, from best to worst:
\begin{enumerate}
\item No-Response, 
\item Random-Response, 
\item One-Step, and 
\item Random.
\end{enumerate}
There appears to be no real benefit to any strategy to going first.
We next extract general trends and observations from the games of each non-random strategy.

\subsection{One-Step Strategy}
One-Step mostly chose {\tt hacking} moves.
When it did not, it chose {\tt cleaning}.
It {\tt cleans} if it has detected a successful attack.
It may {\tt clean} on the first round if it only has low-power attacks (e.g., level 0 or 1).
Third, it may choose to {\tt clean} if many rounds have passed.
This behavior follows from direct application of the evaluation function.
In general, the strategy chooses a move if it is the most likely way to either increase the number of computers it has accounts on, or to decrease the number of computers the opponent has accounts on. Since {\tt cleaning} has a 100\% chance of success in this game, it is an ideal move when another player is known to be on a shared computer. 
Hence, it follows that the strategy will {\tt clean} when it detects a successful attack, but not on failed attacks. 
In the case where no powerful attacks exist, the slight chance that an opponent has started on the same machine as the player means that {\tt cleaning}, which works with 100\% probability, is more likely to improve the net-computers evaluation function than attempting to {\tt hack} another computer.
Similarly, if the prospects for a successful attack are very low (e.g., the best exploit has only power level 1), then {\tt cleaning} is more likely to improve the evaluation function because the knowledge structure models a slight chance that the opponent is on the same machine at the start.
In a third case of using the {\tt clean} action, after several rounds, the strategy has tried enough {\tt hacks} to deduce that the probability of any attack working on the as-yet unexploited machines is very low.
It then follows that the better option is to {\tt clean} the machines it does have access to.

When {\tt hacking}, the highest-power exploits are used first.
If these exploits fail, then other exploits will be attempted.
A common pattern is to attempt to {\tt hack} a computer using the best available exploit, and if it fails, to then {\tt hack} the same computer using the next best available exploit for a different OS.
Probabilistically, this makes sense. 
When a powerful {\tt hack} fails, it provides significant evidence that the OS it assumed was incorrect.
The probability of that powerful attack succeeding on another computer is about 25\% because that computer must have the right OS (a 25\% chance) for it to succeed.
However, if another powerful attack for another OS is available, its probability of success on the same computer is around 33\% because one of the other OSs have already been ruled out.
In contrast, if only one powerful attack is available, the strategy will try that exploit wherever it can.
Only after it has been tried everywhere will the next, in this case weak, exploit be attempted.

This leads to an overall strategy that takes into account the power of the exploits available. 
It can be summarized as follows. 
First, if only weak exploits are available, begin by {\tt cleaning} the machine the player begins on.  
Second, {\tt hack} each machine with the best available exploit. 
On each machine where this fails, {\tt hack} that machine with the next best powerful exploits. 
Also, upon successful detection of a {\tt hack}, {\tt clean} the {\tt hacked} computer.
When the chance of successfully {\tt hacking} additional computers is sufficiently low, proceed defensively and {\tt clean} each controlled computer on each subsequent turn.


\subsection{Random-Response Strategy}
The Random-Response strategy follows many of the same guidelines as does the One-Step strategy regarding when it {\tt cleans} on the first round, how it selects exploits to use in {\tt hacks}, and the fact that it never uses the scouting moves ({\tt recon} and {\tt scan}).
When one powerful exploit is available, the Random-Response strategy, like the One-Step strategy, will continually try that exploit until it has been tried on every computer (excluding the computer it begins the game on). 

One key difference between the Random-Response and the One-Step is that in the Random-Response strategy, a computer is {\tt cleaned} the first turn after it has been succesfully {\tt hacked} into. 
Also, computers are {\tt cleaned} periodically after a certain amount of time. 

These slight changes make Random-Response a slightly more defensive strategy than One-Step.
This difference follows because as the strategy considers its opponents possible moves, it recognizes that there is a possibility it could get {\tt cleaned} off its newly acquired computer.
While this cannot be prevented immediately, a {\tt cleaning} move on the next available turn is the best defense against it.

This slight difference made this strategy more effective against the Random strategy.
Unlike One-Step, which lost when its opponent had a powerful attack, Random-Response managed to win even in those cases.
For example, in one game where the Random strategy had a power level 9 exploit, the Random-Response strategy was able to keep it in check by judicious use of {\tt cleaning} moves, and ended up winning the game.

\subsection{No-Response Strategy}
The No-Response strategy behaved fundamentally differently from the other strategies.
Because it looked ahead at two of its own moves, it was able to value {\tt recon} actions much more highly than the other strategies did.
This follows because a randomly chosen {\tt hack}, even with a powerful exploit, has at most a 25\% chance of success.
As a result, a new computer would be gained approximately every fourth move.
In contrast, a {\tt recon} move reveals the OS of the target machine.
Knowing the OS, the probability of success may be much higher, near one if a good exploit is available for that OS.
So a {\tt recon}-{\tt hack} combination should yield a new computer at approximately every second move, making it the preferable strategy.

When the No-Response strategy applies exploits, it applies whichever exploit is indicated to be effective by the {\tt recon}.
This makes it fundamentally different than the One-Step and Random-Response strategies, which will generally choose the most powerful exploits first.

As the number of viable targets are reduced by exploration and exploitation, the strategy veers toward a defensive posture.
This includes a combination of {\tt backdooring}, {\tt cleaning}, and (occassionally) {\tt scanning}.
The {\tt cleaning} makes sense for the same reasons it was used by One-Step and Random-Response (e.g., after detection of successful attack and when few viable targets remain).
The difference here is that looking the extra step ahead brings into play the relevance of the number of accounts.
A {\tt cleaning} is not guaranteed to remove the opponent from the computer since the opponent may have more accounts than the player.
To know for sure that the opponent is eliminated, one can first {\tt scan} to find out how many accounts there are, and then, if the opponent does not have more accounts than the player has, a {\tt cleaning} will eliminate the opponent.
Additionally, a {\tt backdoor} strengthens the player's hold on the computer and makes a subsequent {\tt clean} move more effective.

A typical game for the No-Response strategy involves a sequence of {\tt recon} moves until a vulnerability is found, and then a successful {\tt hack} move is made.
As the position evolves, the strategy becomes more defensive and combines its ongoing offensive campaign with defensive {\tt backdoor}, {\tt clean}, and  {\tt scan} actions.

\section{Discussion}
\label{sec:discussion}
Our experiments produced both expected and unexpected 
outcomes. It was expected, and the experiments showed, that deeper
tree searches yield better performance. In particular, the Random
strategy, effectively a 0-ply search, performs the worst, as
expected. Still, it manages to defeat the other strategies
occasionally when it starts the game with a powerful
exploit. Furthermore, the 1- and 2-ply searches outperformed the 0-ply
search.  Unfortunately, due to the large breadth of possible moves,
searches of depth three or more were intractable with the available
computational resources; this is discussed below.

Other results were unexpected. Not only did the same evaluation
functions lead to different strategies, but each individual strategy
produced a nuanced variation in tactics depending on the power and
variety of exploits it
begins with. 
In general, more powerful exploits lead to more agressive
play (more attacks), and weaker exploits lead to more defensive play
(more securing of controlled computers). This shows that the
complexity of realistic cyber conflict may not require
complex algorithms. Instead, a simple evaluation function and a
$k$-ply search may suffice to produce highly effective, robust, and
nuanced strategies.

Another unexpected result was the clear dominance of the No-Response
strategy over the Random-Response strategy. This shows that, in this
example of extreme uncertainty about the opponent's situation, it is
more effective to consider one's own moves only. Attempting to average
or search over a wide range of opponent moves when each move has only
a very small probability only seems to help avoid ``knock-out''
moves. (For example, the Random-Response strategy was more likely than
the others to {\tt clean} a newly acquired computer, apparently in
order to avoid being {\tt cleaned} off by the other opponent should
they have happened to be on there first. This observation adds
credibility to the analysis of risk from unilateral actions common in the
literature, such as in attack trees~\cite{schneier1999attack}, attack
graphs~\cite{sheyner2002automated}, and attack Petri
nets~\cite{zakrzewska2011modeling}.

The main shortcoming of this approach is the fact that the
exploration of moves in a $k$-ply search tree is very slow and scales
exponentially with $k$. This made it impossible to explore the
implications of deeper searches. However, a number of methods
for accelerating and deepening the search are available. For example,
in early rounds of the game, there is typically a symmetry among unexplored
computers in the sense that what is known about each of them is the
same. As a result, they lead to the same scores being computed for
moves on each of those computers. 
Rather
than simply compute all of those scores in the tree, the symmetry in
the knowledge can be exploited to drastically reduce the exploration
required. Another approach to speed up the search involves being
more judicious on which branches of the tree to explore. For example,
if a first possible move is especially poor, the tree can be cut at
that point, eliminating all of the computations that belong to its
descendents. These and other heuristics for accelerating computation
have been well studied in the literature~\cite{levy2009computer}. They
have been applied to chess games to provide substantial speed-ups and
to enable much deeper searches since Claude Shannon's seminal work in
1950~\cite{shannon1950programming}.

\section{Conclusions}
\label{sec:conc}

We have shown that an application of game theoretic methods can be used
to analyze a more realistic cyber game than has been heretofore been
studied. In the process, we created strategies that exploit searches
of trees comprised of possible moves. Using simple functions that
evaluate the value of any game situation together with a tree
searching capability, we showed that the resulting strategies can
produce intelligent, nuanced strategies. For example, the resulting
strategies took into account the player's relative advantage based on
the quality of their exploits.  Also, we found that in cases of
extreme uncertainty, it is often better to ignore one's opponent's
possible moves.

One area of interest worthy of further analysis is the investigation
into the value of certainty. 
By creating an alternative
evaluation function that adds some weight to the information content in
the knowledge of the game, the strategy could emphasize an intrinsic
value of learning the OS of computers and the players on each
machine. Such an evaluation function might allow a 1-ply search to
mimic the strategies that were found to be so effective in the 2-ply
No-Response strategy.


%
\bibliographystyle{abbrv}
\bibliography{game}  
%
%

\balancecolumns

\end{document}